\documentclass[12pt]
{article} 
\usepackage{color,setspace,amsmath,amssymb,amsfonts,amsthm}
\usepackage{epsfig}    
\usepackage{graphicx}
\usepackage[ansinew]{inputenc}
\usepackage{times}

\usepackage[
colorlinks=true,backref,pagebackref]{hyperref} 

\def\a{\alpha}

\def\sq2{\sqrt{\frac{\varepsilon_0}{\mu_0}} }

\begin{document}
\title{Comparison of the DeWitt metric in general relativity with the
  fourth-rank constitutive tensors in electrodynamics and in
  elasticity theory} \author{Friedrich
  W. Hehl$^{1,2,}$\footnote{E-mail: hehl@thp.uni-koeln.de}\hspace{5pt}
  and Claus Kiefer$^{1,}$\footnote{E-mail:
    kiefer@thp.uni-koeln.de}\hspace{5pt} \\$^{1}$Institute
  for Theoretical Physics, University of
  Cologne,\\ Z\"ulpicher Str.\ 77, 50937 K\"oln, Germany\\
  $^{2}$Department of Physics and Astronomy, University of Missouri,\\
  Columbia, MO 65211, USA} \date{\begin{footnotesize}\it file
    DWmetric\_13.tex, 07 Dec 2017
  \end{footnotesize}} 
\maketitle

\begin{abstract}We perform a short comparison between the local and
  linear constitutive tensor $\chi^{\lambda\nu\sigma\kappa}$
  in four-dimensional electrodynamics (Sec.2), the elasticity tensor
  $c^{ijkl}$
  in three-dimensional elasticity theory (Sec.3), and the DeWitt
  metric $G^{abcd}$
  in general relativity, with ${a,b,\dots=1,2,3}$
  (Sec.4). We find that the DeWitt metric has only six independent
  components.
\end{abstract}
\begin{footnotesize}{\bf Keywords}: Canonical general relativity,
  DeWitt metric, premetric electrodynamics, elasticity theory,
  constitutive tensors\end{footnotesize}

\section{Introduction}

The discovery of analogies between quantities in different areas of
physics is of great importance. It can help to disclose similar
mathematical structures and to build a conceptual bridge between
physical theories. In this paper, we present such analogies which, to
our knowledge, have largely remained unnoticed. The quantities in
question are the DeWitt metric, which occurs in the canonical
formalism of general relativity, and the constitutive tensors in
electrodynamics and in elasticity theory.

Our paper is organized as follows. In Sec.~2, we review the basic
properties for the constitutive tensor in electrodynamics and in
Sec.~3 we review the properties of the elasticity tensor. The main
part of our paper is Sec.~4. There we discuss analogies of the
DeWitt metric with the tensors discussed in the earlier sections and
point out the important property of the DeWitt metric that it has only
six independent components.  

\section{Four-dimensional electrodynamics} 

The premetric Maxwell equations read \cite{Birkbook,Hehl:2007ut}
$dG=J$ and $dF=0$, where $G$ is the excitation two-form (consisting of
the fields ${\mathbf D}$ and ${\mathbf H}$)\footnote{In
  Ref.\cite{Birkbook}, the excitation is denoted by
  $G=G({\cal D},{\cal H})$.} and $F$ is the field strength two-form
(consisting of the fields ${\mathbf E}$ and ${\mathbf B}$). In
components, these equations read (see \cite{Post:1962}, with
$\mu,\nu,\dots=0,1,2,3$),
\begin{equation}\label{Maxwell1}
  \partial_\nu\mathfrak{G}^{\mu\nu}=\mathfrak{J}^\mu\,,\qquad \partial_\mu
  F_{\nu\lambda}+\partial_\nu F_{\lambda\mu}+ \partial_\lambda
  F_{\mu\nu}=0\,.
\end{equation}
They result from electric charge conservation and magnetic flux
conservation, respectively, and they are independent of any
metric. Thus, only partial derivatives are required in
\eqref{Maxwell1}. Since the fields $\mathfrak{G}^{\mu\nu}$ and
$F_{\mu\nu}$ are unrelated so far, the system in \eqref{Maxwell1} is
not yet able to represent a predictive physical system.

In order to interrelate $\mathfrak{G}^{\mu\nu}$ and $F_{\mu\nu}$, we
take the most general {\it local and linear} constitutive law
\begin{equation}\label{constit1}
  \mathfrak{G}^{\lambda\nu}=\frac
  12\,\chi^{\lambda\nu\sigma\kappa}F_{\sigma\kappa}\,.
\end{equation}
The metric will play a role in this context of the constitutive law
later, see below in \eqref{RiemannVacuum}.  In \eqref{constit1},
$\chi^{\lambda\nu\sigma\kappa}$ is a {\it constitutive tensor density}
of rank 4 and weight $+1$, with the dimension
$[\chi]=[G]/[F]=1/{\rm resistance}$ \cite{Birkbook}.\footnote{See, for
  example, \cite{SV17} for a recent discussion of this tensor.}  Since
both $ \mathfrak{G}^{\lambda\nu}$ and $F_{\sigma\kappa}$ are
antisymmetric in their indices, we have
$\chi^{\lambda\nu\sigma\kappa}=-\chi^{\lambda\nu\kappa\sigma}=
-\chi^{\nu\lambda\sigma\kappa}$.
An antisymmetric pair of indices corresponds, in four dimensions, to
six independent components. Thus, the constitutive tensor can be
considered as a $6\times 6$ matrix with 36 independent components.

A $6\times 6$ matrix can be decomposed in its tracefree symmetric part
(20 independent components), its antisymmetric part (15 components),
and its trace (1~component). At the level of
$\chi^{\lambda\nu\sigma\kappa}$, this {\it decomposition}
under the general linear group $GL(4,R)$ is reflected
in
\begin{eqnarray}\label{chidec}
  \chi^{\lambda\nu\sigma\kappa}&=&\,^{(1)}\chi^{\lambda\nu\sigma\kappa}+
  \,^{(2)}\chi^{\lambda\nu\sigma\kappa}+
  \,^{(3)}\chi^{\lambda\nu\sigma\kappa}\,.\\ \nonumber 36
  &=&\hspace{15pt} 20\hspace{14pt}\oplus \hspace{15pt}15\hspace{14.5pt}
  \oplus \hspace{25pt}1\,.
\end{eqnarray}
The third part, the {\it axion} part \cite{Hehl:2007ut}, is totally
antisymmetric and as such proportional to the Levi-Civita symbol,
$ ^{(3)}\chi^{\lambda\nu\sigma\kappa}:=
\chi^{[\lambda\nu\sigma\kappa]} =\widetilde{\a}\,
\widetilde{\epsilon}^{\lambda\nu\sigma\kappa}$;
note that $\widetilde{\a}$ is a pseudoscalar since
$\widetilde{\epsilon}^{\lambda\nu\sigma\kappa}$ has weight $+1$.
Therefore, the weight of
$\widetilde{\epsilon}^{\lambda\nu\sigma\kappa}$ is essential
information. The second part, the {\it skewon} part, is defined
according to
$ ^{(2)}\chi^{\mu\nu\lambda\rho}:=\frac 12(\chi^{\mu\nu\lambda\rho}-
\chi^{\lambda\rho\mu\nu})$.

If the constitutive equation can be derived from a Lagrangian, which
is the case as long as only {\it reversible} processes are considered,
then $^{(2)}\chi^{\lambda\nu\sigma\kappa}=0$. The {\it principal} part
$^{(1)}\chi^{\lambda\nu\sigma\kappa}$ fulfills the symmetries
$ ^{(1)}\chi^{\lambda\nu\sigma\kappa}=
{}^{(1)}\chi^{\sigma\kappa\lambda\nu}$
and $^{(1)}\chi^{[\lambda\nu\sigma\kappa]}=0$.\footnote{For
  symmetrization and antisymmetrization we use the notation of tensor
  calculus:
  $(\lambda\nu):=\frac{1}{2!}\{\lambda\nu+\nu\lambda\},\,
  [\lambda\nu]:=\frac{1}{2!}\{\lambda\nu-\nu\lambda\}$,
  and corresponding generalizations for p indices, see
  \cite{Schouten:1989}.}  Then, in this case of reversibility,
\begin{eqnarray}\label{chidecREV}
  \chi^{\lambda\nu\sigma\kappa}&=&\,^{(1)}\chi^{\lambda\nu\sigma\kappa}+
  \,^{(3)}\chi^{\lambda\nu\sigma\kappa}\,.\\ \nonumber 21
  &=&\hspace{14.5pt} 20\hspace{15pt}\oplus \hspace{15pt}1\,.
\end{eqnarray}

Note that up to here, we have argued at the level of {\it premetric}
electrodynamics, that is, no metric was involved nor the Lorentz group
$SO(1,3)$. In the special case of a {\it Riemannian} spacetime {\it in
  vacuum,} we have the decomposition ($g:=\det g_{\rho\sigma}$)
\begin{equation}\label{RiemannVacuum}
  \chi^{\lambda\nu\sigma\kappa}=\sqrt{\frac{\varepsilon_0}{\mu_0}}
  \sqrt{-g}\left(g^{\lambda\sigma}g^{\nu\kappa}-g^{\nu\sigma}g^{\lambda\kappa}\right) 
  +\widetilde{\a}\, \widetilde{\epsilon}^{\lambda\nu\sigma\kappa}\,,
\end{equation}
with the electric and the magnetic constants $\varepsilon_0$ and
$\mu_0$, respectively.

\section{Three-dimensional linear elasticity}

In linear elasticity theory for homogeneous bodies, the stress tensor
$\sigma^{ij}=\sigma^{ji}$, with $i,j,\dots=1,2,3$, is related to the
strain tensor
$\varepsilon_{kl}:=\partial_{(k}\,u_{l)}=\varepsilon_{lk}$ by Hooke's
law,
\begin{equation}\label{hooke}
\sigma^{ij}=c^{ijkl}\,\varepsilon_{kl}\,;
\end{equation} 
see, for instance,  \cite{Marsden} or \cite{Sommerfeld}.  
Here, $u_l$ is the displacement
field and $c^{ijkl}$ the constant fourth-rank elasticity tensor.

Since stress and strain are symmetric tensors, the elasticity tensor
obeys the symmetries
\begin{equation}\label{symmetries}
c^{ijkl}=c^{jikl}=c^{ijlk}\,.
\end{equation}
A symmetric second rank tensor has six independent components. Therefore,
by collecting the indices $i$ and $j$ into an index pair and $k$ and
$l$ likewise, the elasticity tensor can be thought of as a $6\times 6$
matrix with 36 independent components.

Usually one assumes that the stress tensor can be derived from an {\it
  elastic potential} $W$ (also called ``strain energy function''),
that is,
\begin{equation}\label{pot}
  \sigma^{ij}=\frac{\partial W}{\partial\varepsilon_{ij}}\,,\qquad{\rm
    or}\qquad c^{ijkl}=\frac{\partial^2\,W}{\partial
    \varepsilon_{ij}\,\partial \varepsilon_{kl}}\,.
\end{equation}
Then the first and the last pair of indices of $c^{ijkl}$ commute,
\begin{equation}\label{paircom}
  c^{ijkl}= c^{klij}\,, 
\end{equation}
and the $6\times 6$ matrix is symmetric and carries only $36-15=21$
independent components. Thus, as is well known, the elasticity tensor
has $21$ independent components.

For the first irreducible piece under the general linear group
$GL(3,R)$, we have
\begin{equation}\label{first}
  ^{(1)}c^{ijkl}:=c^{(ijkl)}\,;
\end{equation}
it has ${3+4-1\choose 4}=15$ independent components. Furthermore,
\begin{equation}\label{rest}
 ^{(2)}c^{ijkl}:=c^{ijkl}-\,^{(1)}c^{ijkl}\,.
\end{equation}
Thus, finally, for the case with an elastic potential, we have the
decomposition
\begin{eqnarray}\label{dec}
 c^{ijkl}&=&\, ^{(1)}c^{ijkl}+\, ^{(2)}c^{ijkl} \,.\\ \nonumber 21
  &=&\hspace{10pt} 15\hspace{11pt}\oplus \hspace{12pt}6\,.
\end{eqnarray}
An analysis by means of the Young tableaux techniques guarantees that
this decomposition is irreducible under the $GL(3,R)$, indeed. 
One
should compare this with the electromagnetic case in
(\ref{chidecREV}).

In this first step of the decomposition of $c^{ijkl}$, we take the
constitutive law \eqref{hooke} at its face value, linking two
symmetric second-rank tensors by a fourth-rank tensor. This tensor can
only be decomposed with respect to $GL(3,R)$. That already such a
first step is physically helpful is shown by the fact that the
vanishing of $^{(2)}c^{ijkl}$ represents the Cauchy relations, which
are fulfilled for certain crystals, see \cite{Haussuhl1,Haussuhl2},
that is, $^{(2)}c^{ijkl}$ plays a role in nature. If,
  additionally, we use the three-dimensional metric of the underlying
space, we can take traces, and a finer decomposition is possible under
the three-dimensional rotation group $SO(3)$. This finer
decomposition, which is, however, not interesting in our
context---since we compare, after all, with the linear decomposition
in (\ref{chidecREV})---has been investigated in detail in
\cite{ItinConf}.

Since $^{(2)}c^{ijkl}$ depends on six independent
  components, one can express it in terms of a symmetric second-rank
  tensor,
\begin{equation}\label{Delta}
\Delta_{mn}:=\frac 14 \epsilon_{mil}\, \epsilon_{njk} \,^{(2)}c^{ijkl}=\Delta_{nm}\,,
\end{equation}
see \cite[Eq.(25)]{Cauchy}.

For {\it isotropic} elastic bodies, with the Lam\'e moduli $\lambda$ and
$\mu$, see \cite{Marsden}, we have
\begin{equation}\label{iso}
  c^{ijkl}=\lambda\,g^{ij}g^{kl}+\mu\left(g^{ik}g^{lj}+g^{il}g^{jk}
  \right)\,.
\end{equation}
Then the decomposition (\ref{dec}) yields \cite{Cauchy},
\begin{eqnarray}\label{c_first}
  ^{(1)} c^{ijkl}&=& (\lambda+2\mu)\,g^{(ij}g^{kl)}\,,\\
\label{c_second}
  ^{(2)} c^{ijkl}&=& \frac{\lambda-\mu}{3}\left(2g^{ij}g^{kl}-
     g^{ik}g^{lj}-g^{il}g^{jk}\right)\,,
\end{eqnarray}
with $g\,\Delta_{ij}= \frac{\lambda-\mu}{2}\,g_{ij}$
  and $g:=\det g_{kl}$.
The piece $^{(2)} c^{ijkl}$ can be called the non-Cauchy part of the
elasticity tensor, see \cite{Itin}, since the conditions
$^{(2)} c^{ijkl}=0$ are called the Cauchy relations of elasticity
theory. The Cauchy relations are only fulfilled if (i) the interaction
forces between the atoms or molecules of a crystal are central forces,
as, for example, in rock salt, (ii) each atom or molecule is a center
of symmetry, and (iii) the interaction forces between the building
blocks of a crystal can be well approximated by a harmonic
potential. In most elastic bodies this is not fulfilled at all, see
\cite{Haussuhl1,Haussuhl2}.

Pure Cauchy materials
$\left(^{(1)}c^{ijkl}\!\ne\! 0,\, ^{(2)}c^{ijkl}\! =\! 0\right)$ and
pure non-Cauchy materials
$\left(^{(1)}c^{ijkl}\!=\! 0,\, ^{(2)}c^{ijkl}\!\ne \! 0\right)$ do
not seem to exist in nature. There are even some plausibility
arguments against pure non-Cauchy materials.

\section{DeWitt metric}

The fundamental dynamical variable in canonical general relativity is
the three-dimensional metric denoted here by $h_{ab}$. As discussed in
detail by DeWitt \cite{DeWitt67}, one can define a metric on the
(unconstrained) configuration space of all such three-metrics; in his
honor, this metric is called DeWitt metric.\footnote{The fundamental
  configuration space of general relativity is {\em superspace}, the
  space of all three-{\em geometries}. The question of a metric on
  superspace is discussed, for example, in
  \cite{DG95,Giulini:2015qha}.} A detailed review can be found in
\cite{Kiefer}.

The DeWitt metric is defined in terms of the inverse three-metric
$h^{ab}$ as follows; see, for example, Eq.(4.25) in
\cite{Kiefer}, 
\begin{equation}\label{WdW}
G^{abcd}=\frac{\sqrt{h}}{2}(h^{ac}h^{bd}+h^{ad}h^{bc}-2h^{ab}h^{cd})\,,
\end{equation}
for $h:=\det h_{ef} \text{ and }a,b,\dots =1,2,3$.  Incidentally,
$G^{abcd}$ is a tensor density of weight +1. We note that
it represents the covariant form of the DeWitt
metric, in spite of the indices to be in the upper position. In the
ADM form of the action, the combination $G^{abcd}K_{ab}K_{cd}$
appears, where $K_{ab}$ are the components of the extrinsic curvature.
 
According to (\ref{WdW}, the DeWitt metric obeys
$G^{abcd}=G^{bacd}=G^{abdc}$ and, moreover, the two index pairs $ab$
and $cd$ are commutative: $G^{abcd}=G^{cdab}$. Therefore, $G^{abcd}$
as well as the constitutive tensor $\chi^{\mu\nu\kappa\lambda}$ and
the elasticity tensor $c^{ijkl}$ are symmetric bilinear forms on a
six-dimensional real vector space. In turn, the irreducible
decomposition of $G^{abcd}$ under the $GL(3,R)$ should be analogous to
(\ref{dec}):
\begin{equation}\label{decWdW}
  G^{abcd}=\,^{(1)}\!G^{abcd}+\,^{(2)}\!G^{abcd}\,,
\end{equation}
with $^{(1)}\!G^{abcd}= G^{(abcd)}$ (15 independent components). If we
take the totally symmetric part of (\ref{WdW}), we recognize that it
vanishes:
\begin{equation}\label{decWdW1}
^{(1)}\!G^{abcd}=0\,.
\end{equation}
Thus,
\begin{equation}\label{decWdW2}
 G^{abcd}=\,
 ^{(2)}\!G^{abcd}=\frac{\sqrt{h}}{2}(h^{ac}h^{bd}+h^{ad}h^{bc}
-2h^{ab}h^{cd})
\end{equation}
carries only {\em six independent components}. 

Clearly, Eq. (\ref{decWdW2}) is similar to (\ref{c_second}), up to a
constant factor and the square root of the metric. In analogy to
elasticity, we can map the DeWitt metric to a symmetric second-rank
tensor according to
\begin{equation}\label{2ndrank}
  \stackrel{G}{\Delta}_{ab}\,:=\frac 14 \epsilon_{akl}\,\epsilon_{bmn}\,
  \frac{1}{\sqrt{h}}  G^{kmnl}=
  \frac{1}{8}\epsilon_{akl}\,\epsilon_{bmn}\left(h^{kn}h^{ml}
    +h^{kl}h^{mn}-2h^{km}h^{nl}\right)\,,
\end{equation}
see \cite{Cauchy,Haussuhl1,Haussuhl2}. The factor
  $1/\sqrt{h}$ was introduced in order to make $G^{abcd}/\sqrt{h}$ a
  tensor density of weight 0, exactly analogous to $^{(2)}c^{ijkl}$ in
  (\ref{Delta}). The second term within the parentheses
of (\ref{2ndrank}) drops out because of its symmetry
in $kl$. If we expand the remaining two terms, we find
\begin{eqnarray}\label{expand}
  \stackrel{G}{\Delta}_{ab}\,&=&\nonumber \frac{1}{8}\left(\epsilon_{akl} \,
    \epsilon_{bmn}h^{kn}h^{ml}  -2\epsilon_{akl} \,
    \epsilon_{bmn}h^{km}h^{nl}\right)=
 \frac{1}{8h}\left(E_{akl} \,
    E_{b}{}^{lk} -2E_{akl} \,
    E_{b}{}^{kl} \right)\\
&=& \frac{1}{4h}\,\left(-h_{ab}-2 h_{ab} \right)
=-\frac{3}{4h}\,h_{ab}\,,
\end{eqnarray}
where the $\epsilon$ denote the Levi-Civita symbols (tensor densities)
and the $E$ the corresponding tensors; see, for example,
\cite{Sokolnikoff}.

Accordingly, the densitized metric $h_{ab}/\sqrt{h}$ can be directly
expressed in terms of the DeWitt metric,
\begin{equation}\label{reciprocal}
  h_{ab}/\sqrt{h}= -\frac 13\, \epsilon_{akl}\,\epsilon_{bmn}\,
  G^{kmnl}\,,
\end{equation}
a truly amazing formula. It is straightforward to show that
(\ref{decWdW2}) and (\ref{reciprocal}) correspond in elasticity to the
isotropic case with, in some suitable units, the Lam\'e constants
$\lambda=-1$ and $\mu=\frac 12$. Thus, the ``compression modulus''
$\cal K$ of the DeWitt metric is ${\cal K}:= \lambda + (2/3)\mu=-2/3$,
a highly unconventional ``material,'' which reacts to pressure with
expansion.

The {\it reciprocity} relation of the DeWitt metric (\cite{Kiefer},
Eq.(4.26)) reads
\begin{equation}\label{reciprocity}
G^{abcd}G_{cdef}^{-1}=\delta^a_{(e}\delta_{f)}^b.
\end{equation}
The inverse DeWitt metric $G_{cdef}^{-1}$ depends on the number of
space dimensions, whereas \eqref{WdW} does not. Furthermore, we
introduced the power to the $-1$ explicitly in order to remind
ourselves that the reciprocal of the DeWitt metric $G_{abcd}^{-1}$
can{\it not} be computed by simply lowering the indices of the DeWitt
metric $G^{abcd}$, see \cite{Giulini:2015qha}.

Let us emphasize that the {\em reciprocal} of the DeWitt metric has
all the (15+6) independent components. The meaning of this should be
investigated. In contrast, in the case of electrodynamics with
$\chi^{\lambda\nu\sigma\kappa}$ and in elasticity with $c^{ijkl}$,
their reciprocals carry the analogous irreducible decompositions as
their originals, that is, they provide no new information.

In canonical general relativity, the relation between the canonical
geometrodynamical momentum $p^{ab}$ and the extrinsic curvature
$K_{cd}$ reads ($G=$ gravitational constant, speed of light $c=1$)
\begin{equation}
\label{momentum}
p^{ab}=\frac{1}{16\pi G}\, G^{abcd}K_{cd}\,,
\end{equation}
see \cite{Kiefer}, Eq.(4.63). This is a Hooke type law of superspace
with the `stress' $p^{ab}$ and the `strain' $K_{cd}$. If such an
interpretation made sense, superspace would constitute a pure
non-Cauchy continuum, that is, a fairly exotic
`substance'. Incidentally, this is reminiscent of Sakharov's
\cite[p.171]{Sakharov:1982} ``metrical elasticity'' of space, here,
however, applied to superspace. It would be like in a crystal, the
elastic constants of which are determined by the underlying molecular
interaction forces.

The DeWitt metric \eqref{WdW} can be generalized to a one-parameter
family of metrics by \cite{DG95,GK94}
\begin{equation}
\label{WdW-alpha}
_{\beta}G^{abcd}=\frac{\sqrt{h}}{2}\left(h^{ac}h^{bd}+h^{ad}h^{bc}
-2\beta h^{ab}h^{cd}\right),
\end{equation}
where $\beta$ is any real number (thus, for general relativity,
$\beta=1$).  If we calculate the total symmetrization of this object,
we find that it is proportional to $1-\beta$, which means that the
case of general relativity is distinguished by having only six
independent components for the DeWitt metric.

Demanding symmetry under permutations of $ab$ and of $cd$,
the inverse of \eqref{WdW-alpha} is unique and can be written in the form
\begin{equation}
\label{WDW-beta}
^{\alpha}\hspace{-1pt}G_{abcd}^{-1}=\frac{1}{2\sqrt{h}}
\left(h_{ac}h_{bd}+h_{ad}h_{bc}-  2\alpha h_{ab}h_{cd}\right),
\end{equation}
where\footnote{In $d$ space dimensions, 
we have $\alpha+\beta=d\alpha\beta$.}
\begin{equation}
\alpha+\beta=3\alpha\beta
\end{equation}
(thus, for general relativity, we have $\alpha=1/2$). Its total
symmetrization is proportional to $1-\alpha$ and consequently is
non-vanishing for the general relativistic case. (It vanishes for the
special case of two space dimensions.)

To summarize, we have disclosed in our paper some interesting
mathematical analogies (and differences) between quantities from
different areas of physics. Whether they point to a deeper physical
connection (e.g. in the context of emergent gravity) remains to be
seen. 

\section*{Acknowledgments}

We thank Masud Chaichian (Helsinki) for a reference to Sakharov's work
in the context of the metrical elasticity of space. Moreover, we thank
Yakov Itin (Jerusalem), Yuri Obukhov (Moscow), and Domenico Giulini
(Hannover) for detailed remarks and corrections to our algebra, and
Nick Kwidzinski and Branislav Nikoli\'c (both of Cologne) for helpful
discussions.


\begin{thebibliography}{99}

\bibitem{Birkbook} F.~W.~Hehl and Yu.~N.~Obukhov, {\it Foundations of
    Classical Electrodynamics: Charge, Flux, and Metric}
  (Birkh\"auser, Boston, MA, 2003).

\bibitem{Hehl:2007ut} F.~W.~Hehl, Y.~N.~Obukhov, J.~P.~Rivera and
  H.~Schmid, Relativistic nature of a magnetoelectric modulus of
    Cr$_2$O$_3$ crystals: A four-dimensional pseudoscalar and its
    measurement, {\em Phys.\ Rev.\ A} {\bf 77}, 022106 (2008); see also
  arXiv:0707.4407 [cond-mat.other].

\bibitem{Post:1962} E.~J.~Post, {\it Formal Structure of
    Electromagnetics -- General Covariance and Electromagnetics}
  (North Holland, Amsterdam, 1962, and Dover, Mineola, NY, 1997).

\bibitem{SV17} S. Schuster and M.~Visser,  	
Effective metrics and a fully covariant description of constitutive
tensors in electrodynamics; see also arXiv:1706.06280 [gr-qc].

\bibitem{Schouten:1989} J.~A.~Schouten, {\it Tensor Analysis for
    Physicists}, 2nd ed. reprinted (Dover, Mineola, NY, 1989).

\bibitem{Marsden} J.~E.~Marsden and T.~J.~R.~Hughes, {\it Mathematical
    Foundations of Elasticity} (Prentice-Hall, Englewoods Cliffs, NJ,
  1983).

\bibitem{Sommerfeld} A.\ Sommerfeld, {\it Mechanik deformierbarer
    Medien,\/} Vorlesungen \"uber Theoretische Physik, Vol.\ II, 5th
  ed. (Akademische Verlagsgesellschaft Geest \& Portig, Leipzig,
  1964).

\bibitem{ItinConf} Y.~Itin and F.~W.~Hehl, Irreducible
    decompositions of the elasticity tensor under the linear and
    orthogonal groups and their physical consequences, {\em J.\ Phys.:
    Conf.\ Ser.} {\bf 597}, 012046 (2015); see also arXiv:1411.5104
    [cond-mat.other].

\bibitem{Cauchy} F.~W.~Hehl and Y.~Itin, The Cauchy relations in
    linear elasticity theory, {\em J.\ Elasticity} {\bf 66}, 185--192
  (2002); see also arXiv:cond-mat/0206175.

\bibitem{Itin} Y.~Itin and F.~W.~Hehl, The constitutive tensor of
    linear elasticity: its decompositions, Cauchy relations, null
    Lagrangians, and wave propagation, {\em J.\ Math.\ Phys.}\ {\bf 54},
  042903 (2013); see also arXiv:1208.1041  [cond-mat.other].

\bibitem{Haussuhl1} S.~Hauss\"uhl, The deviations of the Cauchy
  relations (in German: Die Abwei\-chungen von den Cauchy-Relationen),
  {\em Phys.\ kondens.\ Materie} {\bf 6}, 181--192 (1967).

\bibitem{Haussuhl2} S.~Hauss\"uhl, {\it Physical Properties of
    Crystals: An Introduction} (Wiley-VCH, Weinheim, Germany, 2007).

\bibitem{DeWitt67} B.~S.~DeWitt, Quantum theory of gravity. I.  The
  canonical theory, {\em Phys. Rev.} {\bf 160}, 1113--1148 (1967).

\bibitem{DG95} D.~Giulini, What is the geometry of superspace?  {\em
    Phys. Rev. D} {\bf 51}, 5630--5635 (1995); see also
  arXiv:gr-qc/9311017.

\bibitem{Giulini:2015qha} 
  D.~Giulini,
  Dynamical and Hamiltonian formulation of General Relativity,
  in Chapter 17 of A.~Ashtekar and V.~Petkov (Eds.): {\it Springer
  Handbook of Spacetime} (Springer Verlag, Dordrecht, 2014); see also
  arXiv:1505.01403 [gr-qc].

\bibitem{Kiefer} C.~Kiefer, {\it Quantum Gravity,} 3rd ed. (Oxford
  University Press, Oxford, UK, 2012).

 \bibitem{Sokolnikoff} I.~S.~Sokolnikoff, {\it Tensor
      Analysis} (Wiley, New York, 1951).

\bibitem{Sakharov:1982} A.~D.~Sakharov, {\em Collected Scientific
    Works,} D.~ter~Haar et al.\ (eds.), translated from the Russian
  (Dekker, New York, 1982).

\bibitem{GK94}  D.~Giulini and C.~Kiefer, Wheeler--DeWitt metric and the
           attractivity of gravity, {\em Phys. Lett. A} {\bf 193},
           21--24 (1994); see also  arXiv:gr-qc/9405040.

\end{thebibliography}
\end{document}